\documentclass[%
preprint,
 amsmath,amssymb,
 aps,
]{revtex4-1}

\usepackage{graphicx}% Include figure files
\usepackage{dcolumn}% Align table columns on decimal point
\usepackage{bm}% bold math

\begin{document}

%\preprint{APS/123-QED}

\title{Comparative Study of the Performance of\\
Quantum Annealing and Simulated Annealing}

\author{Hidetoshi Nishimori}
\author{Junichi Tsuda}%
\affiliation{%
Department of Physics, Tokyo Institute of Technology,
Oh-okayama, Meguro-ku, Tokyo 152-8551, Japan
}%

\author{Sergey Knysh}
\affiliation{
QuAIL, NASA Ames Research Center, Moffett Field, CA 94035, USA
}%
\affiliation{
SGT Inc., 7701 Greenbelt Rd, Suite 400, Greenbelt, MD 20770, USA
}% 

\date{\today}
\begin{abstract}
Relations of simulated annealing and quantum annealing are studied by a mapping
from the transition matrix of classical Markovian dynamics of the Ising model
to a quantum Hamiltonian and vice versa. 
It is shown that these two operators, the transition matrix and the Hamiltonian,
share the eigenvalue spectrum.
Thus, if simulated annealing with slow temperature change does not encounter a difficulty
caused by an exponentially long relaxation time at a first-order phase transition, the
same is true for the corresponding process of quantum annealing in the adiabatic limit.
One of the important differences between the classical-to-quantum mapping and the converse
quantum-to-classical mapping is that the Markovian dynamics of a short-range Ising
model is mapped to a short-range quantum system, but the converse mapping from a
short-range quantum system to a classical one results in long-range interactions.
This leads to a difference in efficiencies that simulated annealing
can be efficiently simulated by quantum annealing but the converse is not necessarily true.
We conclude that quantum annealing is easier to implement and is more flexible 
than simulated annealing.
We also point out that the present mapping can be extended to accommodate explicit
time dependence of temperature,  which is used to justify the quantum-mechanical
analysis of simulated annealing by Somma, Batista, and Ortiz.
Additionally, an alternative method to solve the non-equilibrium dynamics of
the one-dimensional Ising model is provided through the classical-to-quantum mapping.

\end{abstract}

\pacs{Valid PACS appear here}% PACS, the Physics and Astronomy
\maketitle

\section{\label{sec:intro}Introduction}

Quantum annealing  has been developed as a generic method to solve combinatorial optimization
problems using quantum-mechanical fluctuations \cite{KN98,KadowakiT,SanT,DC,SIC}.
It is closely related with adiabatic quantum computation \cite{FGGLLP}, which can be
regarded as a restricted version of quantum annealing where the time evolution follows the
adiabatic condition.
Quantum annealing is to be contrasted with simulated annealing, in which classical thermal
fluctuations assist the system to explore the phase space toward the
optimal solution \cite{KSA}.
A large number of comparative studies of quantum annealing and simulated annealing have been
reported from theoretical, numerical, and experimental perspectives, which generally show
superiority of quantum annealing over simulated annealing, at least quantitatively
\cite{KN98,KadowakiT,SanT,DC,MoritaN}.
An early experimental study of a disordered magnet also revealed faster relaxations
toward equilibrium through a quantum path than by a real thermal annealing path \cite{Apl}.
Recent studies concerning the D-Wave machine show mixed results \cite{DW1,DW2,DW3,
DW4,DW5,DW6,DW7,DW8,DW9,DW10,DW11,DW12,DW13,DW14,ARLT}, and further
careful investigations are necessary before firm conclusions are drawn.

The present paper concerns a theoretical analysis to compare
quantum annealing and simulated annealing from a very different viewpoint than the
above-mentioned  studies.
Relations between quantum and classical systems have been known for years
through the path-integral formulation of quantum mechanics \cite{Feynman}
as well as by the Suzuki-Trotter decomposition of the Boltzmann factor
of a quantum system \cite{Suzuki}.
A relatively new development is a mapping of classical Markovian dynamics
to a quantum system, and vice versa, in the same spatial dimension \cite{Henley,Castel}.
This method was originally proposed in the context of the Rokhsar-Kivelson
point of quantum dimer Hamiltonians.
Somma {\it et al.} applied this idea to the analysis of simulated annealing \cite{SBO}
and rederived the
result of Geman and Geman \cite{GG} for the temperature-annealing schedule
through the adiabatic theorem of quantum mechanics.
%See also \cite{BOS} for a related development.
See \cite{BOS} for a related development.
We also refer the reader to a zero-temperature quantum Monte Carlo method employed
in \cite{DC08} for a different type of quantum-to-classical mapping
suitable for classical stochastic studies of quantum systems.

Although the work of Somma {\it et al.} is quite interesting since
it uses quantum mechanics to study a purely classical problem,
it nevertheless includes a few points that need further scrutiny.
First, only the equivalence between the equilibrium state of a
classical system and the ground state of a quantum system has
been emphasized.
However, wider spectra of the transition matrix and the quantum
Hamiltonian should be taken into account to study the detailed behavior
of the energy gap/relaxation time at a quantum/classical phase transition.
Second, the converse mapping from quantum to classical systems needs to be
discussed to complete a comparative study of quantum annealing and simulated annealing, 
in particular to determine whether or not quantum annealing can 
perform a wider class problems than simulated annealing does.
Third, a relation needs to be established between the classical
Markovian dynamics with time-dependent temperature and the time-dependent
Schr\"odinger equation, if we want to know what happens when the
temperature changes relatively quickly
or when quantum annealing is applied beyond the limit of adiabatic evolution.

The goal of the present paper is to  shed new light on the possibilities and
limitations of quantum annealing in comparison with simulated annealing
and to solve the above-mentioned problems.

This paper is organized as follows.
We first review a few basic aspects of Markovian dynamics of the classical Ising
model in Sec. \ref{sec:classical} to fix the notation.
Then, in Sec. \ref{sec:C_to_Q}, we establish a mapping of classical Markovian dynamics
to a quantum Hamiltonian.  A few examples are given for the one-dimensional case.
The converse mapping from quantum to classical systems is given
in Sec. \ref{sec:Q_to_C}.  Similarities and differences between the
classical-to-quantum and quantum-to-classical mappings are discussed.
A more general case of explicitly time-dependent temperature is
analyzed and the work of Somma {\it et al.} is discussed in Sec. \ref{sec:Tt}.
Summary and conclusion are given in the final section.

%%%%%%%%%%%%%%%%%%%%%%%%%%%%%%%%%%
\section{Markovian dynamics of the classical Ising model}\label{sec:classical}

We briefly summarize the Markovian dynamics of the Ising model to fix
the notation.
The temperature $T$, or its inverse $\beta$, is assumed to be time-independent
until otherwise stated at a later section.
The master equation representing the Markovian dynamics is written as
\begin{equation}
\frac{d P_{\sigma}(t)}{dt}=\sum_{\sigma'}W_{\sigma \sigma'}P_{\sigma'}(t)
=\sum_{\sigma'(\ne \sigma)}\big( W_{\sigma\sigma'}P_{\sigma'}(t)
- W_{\sigma'\sigma}P_{\sigma}(t)\big), \label{master_eqn}
\end{equation}
where $\sigma$ is a set of $N$ Ising spins, $\{\sigma_1,\sigma_2,
\cdots, \sigma_N\}$, and $P_{\sigma}(t)$ is the probability that
the system is in the state $\sigma$ at time $t$.
The Hamiltonian of the Ising model will be denoted as $H_0(\sigma)$.
In the context of simulated annealing and quantum annealing, the goal is to find
the ground state of $H_0(\sigma)$.
The transition probability from $\sigma'$ to $\sigma$ is denoted as $W_{\sigma\sigma'}$,
non-vanishing off-diagonal ($\sigma\ne \sigma'$) elements of which satisfy
the detailed balance condition,
\begin{equation}
W_{\sigma\sigma'}P_{\sigma'}^{(0)}=W_{\sigma'\sigma}P_{\sigma}^{(0)}\quad
\Big(P_{\sigma}^{(0)}=\frac{e^{-\beta H_0(\sigma)}}{Z},
~Z=\sum_{\sigma}e^{-\beta H_0(\sigma)}\Big).
\label{db0}
\end{equation}
We write $\hat{W}$ for the $2^N\times 2^N$ matrix with elements
$(\hat{W})_{\sigma\sigma'}=W_{\sigma\sigma'}$.
All matrices and vectors will be represented in the $\sigma$-basis.
The non-vanishing off-diagonal element of $\hat{W}$ can be expressed as
\begin{equation}
W_{\sigma\sigma'}=w_{\sigma\sigma'}
e^{-\frac{1}{2}\beta (H_0(\sigma)-H_0(\sigma'))},
\end{equation}
where $w_{\sigma\sigma'}$ is symmetric, $w_{\sigma\sigma'}=w_{\sigma'\sigma}$, 
according to the detailed balance condition (\ref{db0}).

The quantity $w_{\sigma\sigma'}$ can be chosen arbitrarily as long as
the resulting $W_{\sigma\sigma'}$ can be regarded as a conditional probability.
For example, the Metropolis update rule has
\begin{equation}
w_{\sigma\sigma'}=\min\big(e^{-\frac{1}{2}\beta (H_0(\sigma')-H_0(\sigma))},
e^{\frac{1}{2}\beta (H_0(\sigma')-H_0(\sigma))}\big),
\label{metropolis}
\end{equation}
and  the heat-bath method is realized by
\begin{equation}
w_{\sigma\sigma'}=\frac{1}{\displaystyle
e^{-\frac{1}{2}\beta (H_0(\sigma')-H_0(\sigma))}+e^{\frac{1}{2}\beta (H_0(\sigma')-H_0(\sigma))}}.
\label{heat-bath-w}
\end{equation}

The eigenvalues of the transition matrix $\hat{W}$ are negative semi-definite.
The largest eigenvalue is 0 and corresponds to thermal equilibrium.
If we denote the eigenvalues as
$\lambda_0=0>\lambda_1>\lambda_2>\cdots$,
a general solution to the master equation (\ref{master_eqn}) is written as
\begin{equation}
P_{\sigma}(t)=\sum_{n=0}a_n e^{-|\lambda_n|t}\psi_{\sigma}^{(R,n)}.
\label{solution_master_eq}
\end{equation}
Here, $\psi_{\sigma}^{(R,n)}$ is the $\sigma$ component of the
$n$th right eigenvector $\hat{\psi}^{(R,n)}$ of $\hat{W}$,
\begin{equation}
\hat{W}\hat{\psi}^{(R,n)}=\lambda_n\hat{\psi}^{(R,n)}.
\end{equation}
In particular, the right eigenvector corresponding to $\lambda_0=0$ is
\begin{equation}
\psi_{\sigma}^{(R,0)}=P_{\sigma}^{(0)}. \label{P0}
\end{equation}

%%%%%%%%%%%%%%%%%%%%%%%%%%%%%%%
\section{Quantum Hamiltonian derived from classical dynamics}\label{sec:C_to_Q}

We now derive a quantum Hamiltonian $\hat{H}$ from the classical transition matrix $\hat{W}$.
The original idea comes from Castelnovo {\it et al.} \cite{Castel}, but we proceed with
carefully keeping in mind the correspondence between quantum annealing and simulated annealing.

\subsection{Construction of quantum Hamiltonian}

Let us denote by $\hat{H}_0$  the diagonal matrix with $H_0(\sigma)$
as its diagonal elements, $(\hat{H})_{\sigma\sigma}=H_0(\sigma)$.
A quantum Hamiltonian is then defined from $\hat{W}$ as
\begin{equation}
\hat{H}=-e^{\frac{1}{2}\beta \hat{H}_0}\hat{W}e^{-\frac{1}{2}\beta \hat{H}_0}.
\label{C_to_Q}
\end{equation}
It is straightforward to verify that $\hat{H}$ is real and symmetric, {\it i.e.} Hermitian,
using the detailed balance condition (\ref{db0}).
We can therefore regard $\hat{H}$ as the Hamiltonian of a quantum system.
The eigenvalue spectrum of $\hat{W}$,
\begin{equation}
\hat{W}\hat{\psi}^{(R,n)}=\lambda_n \hat{\psi}^{(R,n)}, \label{W_spectrum}
\end{equation}
is shared with $\hat{H}$:
\begin{align}
\hat{H}\hat{\phi}^{(n)}
=-e^{\frac{1}{2}\beta\hat{H}_0}\hat{W}\hat{\psi}^{(R,n)}
=-\lambda_n \hat{\phi}^{(n)},  \label{eigenvalues}
\end{align}
where
\begin{equation}
\hat{\phi}^{(n)}=e^{\frac{1}{2}\beta \hat{H}_0} \hat{\psi}^{(R,n)}. \label{eigenvector}
\end{equation}
Equations (\ref{W_spectrum})-(\ref{eigenvector}) show one-to-one correpondence between
the eigenvalues and eigenvectors of $\hat{H}$ and $\hat{W}$, which establishes a
classical-to-quantum mapping in the same spatial dimension.

The classical Ising model $\hat{H}_0$ has  the relaxation time toward equilibrium 
as $\tau=1/|\lambda_1|$ according to Eq. (\ref{solution_master_eq}).
If $\hat{H}_0$ has a phase transition at a temperature $T_{\rm c}$,
the relaxation time diverges at $T_{\rm c}$ as a function of the system size $N$.  
If the transition is of second order, $\tau$ diverges polynomially $\tau \propto N^a~(a>0)$,
and the divergence is exponential $\tau \propto e^{bN}~(b>0)$ at a first-order transition.
Correspondingly, the quantum system $\hat{H}$ has a quantum phase transition
at the system parameter determined by the correspondence (\ref{C_to_Q}).
The energy gap $\Delta =|\lambda_1|$ between the ground state (whose energy is $\lambda_0=0$)
and the first excited state closes polynomially $\Delta \propto N^{-a}$ at a 
second-order transition and exponentially $\Delta \propto e^{-bN}$ at a first-order transition.

It should be kept in mind that these discussions apply to the case of time-independent
temperature for the classical dynamics and stationary states for the quantum system.
This means that,  in the context of simulated annealing, the system is supposed to
evolve in quasi-equilibrium, {\it i.e.} the temperature changes very slowly such that
the system stays very close to thermal equilibrium.
The corresponding quantum system is  driven adiabatically, 
and the system is kept infinitesimally close to the instantaneous stationary state.
The case with strong time dependence of temperature in simulated annealing
and non-adiabatic evolution in quantum annealing will be analyzed in Sec. \ref{sec:Tt}.

The normalized ground-state wave function of  $\hat{H}$ is written as
\begin{equation}
\hat{\phi}^{(0)}=\frac{e^{-\frac{1}{2}\beta \hat{H}_0}}{\sqrt{Z}}\sum_{\sigma}|\sigma\rangle,
\label{phi_0}
\end{equation}
according to Eqs. (\ref{P0}) and (\ref{eigenvector}).
If we write $\langle \hat{Q} \rangle_{0}$ for the expectation value of a matrix $\hat{Q}$
diagonal in the $\sigma$-basis by the ground-state 
wave function (\ref{phi_0}), this expectation value is equal to the thermal
expectation value of the corresponding classical system,
\begin{equation}
\langle \hat{Q} \rangle_{0}=\frac{1}{Z}\sum_{\sigma}\langle \sigma |\hat{Q}|\sigma\rangle
e^{-\beta H_0(\sigma)}.
\end{equation}

\subsection{Explicit formulas for the quantum Hamiltonian}

We next derive the explicit form of
$\hat{H}$.  Non-vanishing off-diagonal elements are
\begin{equation}
(\hat{H})_{\sigma\sigma'}=H_{\sigma\sigma'}=-e^{\frac{1}{2}\beta H_0(\sigma)}W_{\sigma\sigma'}
e^{-\frac{1}{2}\beta H_0(\sigma')}
=-w_{\sigma\sigma'}~(<0).
\end{equation}
Diagonal elements are
\begin{equation}
H_{\sigma\sigma}=-W_{\sigma\sigma}
=\sum_{\sigma'(\ne \sigma)} W_{\sigma'\sigma}
=\sum_{\sigma'(\ne \sigma)}w_{\sigma'\sigma}e^{-\frac{1}{2}\beta
(H_0(\sigma')-H_0(\sigma))},
\end{equation}
where the condition of probability conservation, $\sum_{\sigma'}W_{\sigma'\sigma}=0$,
has been used.
These equations lead to the following form of
$\hat{H}$, using $w_{\sigma\sigma'}=w_{\sigma'\sigma}$,
\begin{align}
\hat{H}&=\frac{1}{2}\sum_{\sigma\sigma'}w_{\sigma\sigma'}
\big(e^{-\frac{1}{2}\beta (H_0(\sigma')-H_0(\sigma))}
|\sigma\rangle \langle \sigma |
+e^{\frac{1}{2}\beta (H_0(\sigma')-H_0(\sigma))}
|\sigma'\rangle \langle \sigma' |
-|\sigma'\rangle \langle \sigma|-|\sigma\rangle \langle \sigma'|\big)\\
&=\sum_{\sigma}\sum_{\sigma'} w_{\sigma\sigma'} \big(e^{-\frac{1}{2}\beta (H_0(\sigma')-H_0(\sigma))}
|\sigma\rangle \langle \sigma| -|\sigma'\rangle \langle \sigma|\big).
\label{H_hat}
\end{align}
The second term of this last expression represents a transverse-field term if
$\sigma'$ is different from $\sigma$ only by a single-spin flip because the
transverse-field operator $\sigma_i^x$ flips a single spin at site $i$.
The first term is then a diagonal interaction of a usual classical Ising model
with interaction range comparable to that of the original classical Ising model
because the quantity in the exponent, $H_0(\sigma')-H_0(\sigma)$,
includes only local interactions if $\sigma'$ and $\sigma$ are different
at a single site.
Examples will be given below.
It has hence been shown that Markovian dynamics of a classical Ising model
with short-range interactions is equivalent to the stationary-state
quantum mechanics of a transverse-field Ising model with comparable interaction range.
It is concluded that simulated annealing under quasi-static condition can be exactly
mapped to quantum annealing under adiabatic condition.
In other words, if a given combinatorial optimization problem expressed in terms
of a short-range Ising model can be solved efficiently by simulated annealing
in the sense that no problematic first-order phase transition occurs in the process,
the same is always possible by quantum annealing.
In this sense, the efficiency of quantum annealing is at least comparable
to that of simulated annealing.

\subsection{One-dimensional Ising model}

As a concrete example, let us discuss the simple case of the one-dimensional
Ising model with nearest-neighbor interactions under a periodic boundary condition.
The dynamics is supposed to proceed under single-spin flip processes.
Since $\sigma'$ is different from $\sigma$ only at a site,
which is chosen as site $j$,
\begin{equation}
H_0(\sigma')-H_0(\sigma)=2J\sigma_j (\sigma_{j-1}+\sigma_{j+1})=-2H_j,
\label{H0_diff}
\end{equation}
where the final equality defines $H_j$.

First, for the heat-bath dynamics with Eq. (\ref{heat-bath-w}),
the diagonal and off-diagonal coefficients in Eq. (\ref{H_hat}) are
\begin{align}
w_{\sigma\sigma'}e^{-\frac{1}{2}\beta (H_0(\sigma')-H_0(\sigma))}
=\frac{e^{\beta H_j}}{e^{\beta H_j}+e^{-\beta H_j}},\quad
w_{\sigma\sigma'}=\frac{1}{e^{\beta H_j}+e^{-\beta H_j}}.
\end{align}
It is relatively straightforward to evaluate these expressions using
Eq. (\ref{H0_diff}) to find the following  formula of the
quantum Hamiltonian,
\begin{equation}
\hat{H}=\frac{N}{2}-\frac{1}{2}\tanh 2K 
\sum_{j=1}^N \sigma_j^z \sigma_{j+1}^z
-\frac{1}{2\cosh 2K}\sum_{j=1}^N
\big(\cosh^2K -\sinh^2 K \, \sigma_{j-1}^z \sigma_{j+1}^z\big)\sigma_j^x,
\label{hhat-hb}
\end{equation}
where $K=\beta J$, and $\sigma_j$ has been replaced by the Pauli matrix $\sigma_j^z$.
Equation (\ref{hhat-hb}) is a one-dimensional transverse-field Ising model
with nearest-neighbor interactions.
In the high-temperature limit $K=0$, Eq. (\ref{hhat-hb}) reduces to a non-interacting
transverse-field Hamiltonian,
\begin{equation}
\hat{H}=\frac{N}{2}-\frac{1}{2}\sum_{j=1}^N\sigma_j^x, \label{K0}
\end{equation}
whose ground state is completely disordered in the $\sigma^z$-basis.
This is exactly the initial state of quantum annealing.
In the opposite limit $K\to \infty$, 
\begin{equation}
\hat{H}=\frac{N}{2}-\frac{1}{2}\sum_{j=1}^N \sigma_j^z\sigma_{j+1}^z
-\frac{1}{4}\sum_{j=1}^N \big(1-\sigma_{j-1}^z\sigma_{j+1}^z\big)\sigma_j^x. \label{Kinf}
\end{equation}
The state with all $\sigma_j^z$ having eigenvalue 1 is an eigenstate
of this Hamiltonian.  
The Perron-Frobenius theorem assures that this is the unique ground state.
Thus, the quasi-static simulated annealing from  high temperature to zero temperature
has been mapped to the behavior of the quantum system starting from the
 disordered state and ending up in the  ordered state
after an adiabatic evolution.

The usual transverse-field Ising model with the Hamiltonian
\begin{equation}
\hat{H}=-J\sum_{j}\sigma_j^z \sigma_{j+1}^z -\Gamma \sum_j \sigma_j^x
\end{equation}
has a phase transition at $\Gamma /J=1$.
In contrast, the present model (\ref{hhat-hb}) with the additional term involving
$\sigma_{j-1}^z\sigma_{j+1}^z$ in front of $\sigma_j^x$
has no phase transition between the two limiting cases of
Eqs. (\ref{K0}) and (\ref{Kinf}) because the original
classical Ising model has no finite-temperature transition.
We thus conclude that the additional term in Eq. (\ref{hhat-hb}) having
$\sigma_{j-1}^z\sigma_{j+1}^z$  drives the system away from the quantum critical point,
thus realizing a smooth (non-singular) process in the course of quantum annealing.
In Eq. (\ref{hhat-hb}), the coefficient of the transverse-field term is small
($\cosh^2K-\sinh^2K=1$) when the local spin alignment is ferromagnetic
$\sigma_{j-1}^z\sigma_{j+1}^z=1$ and is large ($\cosh^2K+\sinh^2K>1$) when the spin
alignment is different from the target state $\sigma_{j-1}^z\sigma_{j+1}^z=-1$.
This means that the local, adaptive change of the coefficient
of transverse field is effective to avoid problematic
quantum phase transitions in quantum annealing.
Although this lesson has been extracted from the simple one-dimensional
Ising model, it may be worth considering to implement a similar process
of adaptive change of the coefficient of the quantum driving term
in more complicated cases when one encounters difficulties in quantum annealing.

Another comment concerns the exact solution of the quantum system (\ref{hhat-hb}).
This Hamiltonian can be diagonalized by the Jordan-Wigner transformation
as will be discussed in the next section.
This serves as an additional route to the complete solution of
the  dynamics of the one-dimensional classical Ising model
pioneered by Glauber \cite{Glauber}.

The Metropolis method with Eq. (\ref{metropolis}) can be analyzed in the same manner.
The resulting quantum Hamiltonian is
\begin{align}
\hat{H}&=\frac{N}{4}(3+e^{-4K})\nonumber\\
&-\frac{1}{4}(1-e^{-4K})
\sum_{j=1}^N \big(2\sigma_j^z\sigma_{j+1}^z+\sigma_{j-1}^z\sigma_{j+1}^z\big)
-\frac{1}{2}(1+e^{-2K})\sum_{j=1}^N
 \big(1-\tanh K\, \sigma_{j-1}^z\sigma_{j+1}^z\big)\sigma_j^x.
 \label{hhat-Metropolis}
\end{align}
We again find that the coefficient of the transverse-field term
is adaptively changed according to the alignment
of the local spins, $\sigma_{j-1}^z\sigma_{j+1}^z$.
Notice that the diagonal interaction term now involves next-nearest-neighbor
interactions.  
It is of course still of short-range, but this example shows
that the range generally changes slightly.

It is also possible to implement random interactions,
\begin{equation}
H_0(\sigma)=-\sum_{j=1}^N J_j \sigma_{j-1}\sigma_j.
\end{equation}
The final expression of the Hamiltonian for the heat-bath update rule is then
\begin{align}
\hat{H}&=\frac{N}{2}-\frac{1}{2}\sum_j \frac{c_j s_j}{c_j^2 c_{j+1}^2-s_j^2 s_{j+1}^2}
\sigma_{j-1}^z \sigma_{j}^z
-\frac{1}{2}\sum_j \frac{c_{j+1}s_{j+1}}{c_j^2 c_{j+1}^2-s_j^2 s_{j+1}^2}\sigma_j^z \sigma_{j+1}^z\\
&-\frac{1}{2}\sum_j \left(\frac{c_jc_{j+1}}{c_j^2 c_{j+1}^2 -s_j^2 s_{j+1}^2}
-\frac{s_j s_{j+1}}{c_j^2 c_{j+1}^2-s_j^2 s_{j+1}^2}\sigma_{j-1}^z \sigma_{j+1}^z\right)\sigma_j^x,
\end{align}
where $c_j=\cosh \beta J_j$ and $s_j=\sinh \beta J_j$.
This Hamiltonian can  be reduced to a quadratic form of Fermion
by the Jordan-Wigner transformation.
It is not possible to completely diagonalize the quadratic
form using Fourier transformation due to the lack of translational invariance.
The quadratic expression nevertheless would give us a tool to analyze the
classical dynamics of the one-dimensional disordered Ising model 
by numerical diagonalization of large systems.

\subsection{Non-equilibrium dynamics of the one-dimensional Ising model}

The quantum Hamiltonian of Eq. (\ref{hhat-hb}) representing
the heat-bath dynamics of the one-dimensional Ising model can be solved exactly by
an application of the Jordan-Wigner transformation.
Before it is applied, we transform the Hamiltonian to a more customary form by
performing $\pi/2$ rotations about the $y$-axis so that $x \to z$ and $z \to - x$.
Following this transformation, local fields are along the $z$-direction, coupling
to $\sigma_j^z$, two-body interactions are proportional to $\sigma_{j-1}^x \sigma_j^x$,
and the three-body terms are $\propto \sigma_{j-1}^x \sigma_j^z \sigma_{j+1}^x$.

We introduce new operators
\begin{equation}
  a_j = \frac{\sigma_j^x - i \sigma_j^y}{2}  \prod_{\ell = 1}^{j - 1}
  \left( - \sigma_{\ell}^z \right) \quad \textrm{and} \quad
  a_j^\dag = \frac{\sigma_j^x + i \sigma_j^y}{2}  \prod_{\ell = 1}^{j -
  1} \left( - \sigma_{\ell}^z \right),
\label{aadag}
\end{equation}
which can be verified to obey the Fermionic anti-commutation relations: 
$\bigl\{ a_j, a_k^\dag \bigr\} = \delta_{j k}$ and 
$\{ a_j, a_k \} = \bigl\{ a_j^\dag, a_k^\dag \bigr\} = 0$. With this substitution, the
Hamiltonian (\ref{hhat-hb}) may be rewritten as
\begin{equation}
  \hat H = C + J_1 \sum_{j=1}^N \bigl( a_j - a_j^\dag \bigr) \bigl( a_{j+1} + a_{j+1}^\dag \bigr) 
  + J_2 \sum_{j=1}^N \bigl( a_{j-1} - a_{j-1}^\dag \bigr) \bigl( a_{j+1} + a_{j+1}^\dag \bigr) 
  - \Gamma \sum_{j=1}^N \bigl( a_j^\dag a_j - a_j a_j^\dag \bigr),
  \label{Haa}
\end{equation}
with $C =N/2$, $J_1 =(\tanh 2K)/2$, $J_2 =
\sinh^2 K/(2 \cosh 2 K)$, and $\Gamma = \cosh^2 K/(2 \cosh 2 K)$.
Interestingly, because Fermionic annihilation and creation operators
(\ref{aadag}) carry a chain product $\prod_\ell (-\sigma_\ell^z)$, the
three-body terms of the form $\sigma_{j-1}^x \sigma_j^z \sigma_{j+1}^x$
become quadratic after the transformation.
For the Metropolis dynamics, the quantum Hamiltonian Eq. (\ref{hhat-Metropolis})
contains also terms  $\propto \sigma_{j-1}^x \sigma_{j+1}^x$ giving rise to quartic terms.
The exact analytical solution is possible only for the heat-bath update rule.

Because of the way the boundary terms $\sigma_N^x \sigma_1^x$, 
$\sigma_N^x \sigma_1^z \sigma_2^x$, and $\sigma_{N-1}^x \sigma_N^z \sigma_1^x$ are treated
in applying the transformation (\ref{aadag}), boundary conditions require special treatments.
For states with even number of Fermions, anti-periodic
boundary conditions ($a_{N+k} \equiv -a_k$) should be used in Eq.~(\ref{Haa}); 
periodic boundary conditions ($a_{N+k} \equiv a_k$) will be used for states with odd number
of Fermions.

Diagonalization of Eq. (\ref{Haa}) is performed using a variant of the Bogolyubov transformation.
The quadratic form Eq. (\ref{Haa}) can be written in a matrix form as
\begin{equation}
\left( \begin{matrix} \mathbf a^\dag & \mathbf a \end{matrix} \right)
\left( \begin{matrix} \mathbf A & \mathbf B \\ 
  -\mathbf B & -\mathbf A \end{matrix} \right)
\left( \begin{matrix} \mathbf a \\ \mathbf a^\dag \end{matrix} \right),
\label{Hblock}
\end{equation}
where the only non-zero elements of
matrices $\mathbf A$ and $\mathbf B$ are 
\begin{gather}
  A_{j,j}=-\Gamma, \quad A_{j,j\pm1}=-\tfrac{1}{2}J_1, \quad
  A_{j,j\pm2}=-\tfrac{1}{2}J_2, \label{Amat} \\
  B_{j,j\pm1}=\mp\tfrac{1}{2}J_1, \quad B_{j,j\pm2}=\mp\tfrac{1}{2}J_2. \label{Bmat}
\end{gather}
Here we assume that matrix indices are periodic (e.g. $B_{1,-1} \equiv B_{1,N-1}$).
Matrix (\ref{Hblock}) can be diagonalized in terms of new quasiparticles with
annihilation/creation operators $\gamma_j$, $\gamma_j^\dag$ connected
to $a_j$, $a_j^\dag$ via a linear transformation
\begin{equation}
\left( \begin{matrix} \mathbf a \\ \mathbf a^\dag \end{matrix} \right)
= \left( \begin{matrix} \mathbf U & \mathbf V \\ 
\mathbf V & \mathbf U \end{matrix} \right)
\left( \begin{matrix} \boldsymbol\gamma \\ 
\boldsymbol\gamma^\dag \end{matrix} \right),
\end{equation}
so that
\begin{equation}
  \hat H = \bigl( \begin{matrix} \boldsymbol\gamma^\dag & 
    \boldsymbol\gamma \end{matrix} \bigr)
  \Biggl( \begin{matrix} \boldsymbol\epsilon & \mathbf 0 \\ 
    \mathbf 0 & -\boldsymbol\epsilon \end{matrix} \Biggr)
  \Biggl( \begin{matrix} \boldsymbol\gamma \\ 
    \boldsymbol\gamma^\dag \end{matrix} \Biggr)
  \equiv \sum_\alpha \epsilon_\alpha 
  \bigl( \gamma_\alpha^\dag \gamma_\alpha - 
  \gamma_\alpha \gamma_\alpha^\dag \bigr)
\end{equation}
Lastly, we perform another transformation
\begin{equation}
\left( \begin{matrix} \mathbf F \\ \mathbf G \end{matrix} \right)
=\frac{1}{\sqrt{2}} \left( \begin{matrix} \mathbf 1 & \mathbf 1 \\ 
-\mathbf 1 & \mathbf 1 \end{matrix} \right)
\left( \begin{matrix} \mathbf U \\ \mathbf V \end{matrix} \right)
\end{equation}
to obtain a particularly compact formulation.
Single-particle energies corresponding to diagonal elements of $\boldsymbol\epsilon$
satisfy the eigenvalue equation,
\begin{equation}
\begin{split}
  \Gamma f_j + J_1 f_{j + 1} + J_2 f_{j + 2} &= \epsilon g_j \\
  \Gamma g_j + J_1 g_{j - 1} + J_2 g_{j - 2} &= \epsilon f_j ,
  \label{eqfg}
\end{split}
\end{equation}
where $f_j$ and $g_j$ are, respectively, columns of $\mathbf F$ and $\mathbf G$.
Solutions to Eq. \eqref{eqfg} can be sought in the form 
$f_j=fe^{i p j}, g_j=g e^{i p j}$,
where $p$ is the momentum: $p = \pi (2k+1)/N$ for a sector
with even number of Fermions and $p = 2\pi k/N$ for the odd sector
($k = 0, 1, \ldots, N - 1$).
From the vanishing condition of the determinant for the system above, we obtain
\begin{equation}
  \epsilon_p^2 = \left| \Gamma + J_1 e^{i p} 
  + J_2  \mathrm{e}^{2 \mathrm{i} p} \right|^2 
  = \frac{1}{4} \left( 1 + \tanh 2 K \cdot \cos p \right)^2.
  \label{ep}
\end{equation}

No single-particle states with positive energies are occupied in the ground state.
The ground-state energy is $E_0=C-\sum_p \epsilon_p$ which is
trivially verified to be zero as should be expected. Energies of
excited states can be written as
\begin{equation}
  E_{p_1,\ldots,p_\nu} = 2 \epsilon_{p_1} + 2 \epsilon_{p_2} + \cdots 
  + 2 \epsilon_{p_\nu},
  \label{Ep}
\end{equation}
corresponding to $\nu$ excitations with momenta $p_1,\ldots,p_\nu$
chosen from the appropriate set, depending on the parity of $\nu$.
This additive form is in general agreement with the original analysis
by Glauber who used a different technique to find the spectrum \cite{Glauber}.
In the more general case of random interactions, nearest-neighbor and
next-nearest-neighbor couplings become site-dependent. Single-particle
energies are easily obtained numerically by diagonalizing a sparse matrix.

From Eqs.~(\ref{ep}) and (\ref{Ep}) we see that the gap remains finite
in the thermodynamic limit $\Delta =E_{\textrm{min}} = 1-\tanh 2K>0$ at non-zero
temperature, consistent with a lack of phase transition for the classical model.

%%%%%%%%%%%%%%%%%%%%%%%%%%%%5
\section{Quantum Hamiltonian to classical dynamics}\label{sec:Q_to_C}

The next step is to find a converse mapping from a quantum Hamiltonian to
classical dynamics, again following Castelnovo {\it et al} \cite{Castel}.

Suppose we are given a quantum Hamiltonian $\hat{H}$, whose ground-state energy
is chosen to be 0 by a shift of the energy standard, $\hat{H}\hat{\phi}^{(0)}=0$.
In order to derive the Markovian dynamics of a classical Ising model from
the quantum Hamiltonian $\hat{H}$, we assume that this $\hat{H}$ is
represented in the basis to diagonalize $\{\sigma_i^z\}_i$ and also that off-diagonal
elements are negative semi-definite, $H_{\sigma\sigma'}\le 0~(\sigma\ne\sigma')$.
Then, according to the Perron-Frobenius theorem applied to $\hat{H}'=-\hat{H}$, 
the eigenvector $\hat{\phi}^{(0)}$ of $\hat{H}'$ for the largest eigenvalue is not
degenerate and all its elements can be chosen to be positive.
This allows us to take the logarithm of each element to define the classical Ising model,
\begin{equation}
H_0(\sigma)=-2\log \phi_\sigma^{(0)}.\label{q-to-c_H0}
\end{equation}
This definition is motivated by the opposite mapping (\ref{phi_0}) up to a constant.
Then, the matrix defined by
\begin{equation}
\hat{W}=-e^{-\frac{1}{2} \hat{H}_0}\hat{H}e^{\frac{1}{2} \hat{H}_0}
\label{W_from_H0}
\end{equation}
satisfies the following conditions required for a transition
matrix of classical dynamics,
\begin{align}
W_{\sigma\sigma'}&\ge 0\quad (\sigma\ne \sigma') \label{Wss}\\
 (1,1,1,\cdots ,1)\hat{W}&=0 \label{consv}\\
 \hat{W} e^{-\beta \hat{H}_0}\sum_{\sigma}|\sigma \rangle &=0  \label{equil}\\
 W_{\sigma\sigma'}e^{-H_0(\sigma')}&=W_{\sigma'\sigma}e^{-H_0(\sigma)}. \label{db}
\end{align}
Equation (\ref{Wss}) follows from $H_{\sigma\sigma'}\le 0$.
Equation (\ref{consv}) for the conservation of probability comes from 
\begin{equation}
\sum_{\sigma}W_{\sigma\sigma'}=-\sum_{\sigma}e^{-\frac{1}{2}H_0(\sigma)}
H_{\sigma\sigma'}e^{\frac{1}{2}H_0(\sigma')}
=-\sum_{\sigma}\phi_{\sigma}^{(0)}H_{\sigma\sigma'}e^{\frac{1}{2}H_0(\sigma')}
=0,
\end{equation}
where we have used $\hat{H}\hat{\phi}^{(0)}=0$.
Equation (\ref{equil}) for equilibrium is due to $\hat{H}\hat{\phi}^{(0)}=0$.
Finally, Eq. (\ref{db}) can be derived from Eq. (\ref{W_from_H0}).

A quantum-to-classical mapping has thus been established.
An important difference from the opposite classical-to-quantum mapping
is the range of interactions in the resulting classical Hamiltonian.
To accommodate the values of $\phi_{\sigma}^{(0)}$ for all spin
configurations of $\sigma =(\sigma_1,\cdots, \sigma_N)$, the
Hamiltonian $H_0(\sigma)$ of Eq. (\ref{q-to-c_H0}) should be expressed as a linear
combination of all possible products and sums of spin variables,
\begin{equation}
H_0(\sigma)=J^{(0)}+\sum_i J_i^{(1)}\sigma_i
+\sum_{i,j}J_{ij}^{(2)}\sigma_i\sigma_j+\sum_{ijk}J_{ijk}^{(3)}
\sigma_i \sigma_j\sigma_k +\cdots +J^{(N)}\sigma_1 \sigma_2\cdots \sigma_N.
\label{H_0_q-to-c}
\end{equation}
By relating this expression with $\phi_{\sigma}^{(0)}$ following Eq. (\ref{q-to-c_H0})
and assigning all possible values of $\sigma$ to Eq. (\ref{q-to-c_H0}), we obtain a set
of linear equations for the $2^N$ coefficients $J^{(0)}, \{J_i^{(1)}\}_i, \cdots, J^{(N)}$.
Its solution generally has non-vanishing values of all those coefficients.
This means that the Hamiltonian $H_0$ has very complicated multibody
long-range interactions as given in Eq. (\ref{H_0_q-to-c}) even if the original quantum
Hamiltonian $\hat{H}$ has only short-range interactions.
Although the eigenvalues and eigenstates are shared by the quantum $\hat{H}$ and the
classical $\hat{W}$, an implementation of the classical dynamics in simulated annealing
is actually inefficient due to the complicated interactions.
This is in marked contrast with the opposite classical-to-quantum
mapping, where short-range interactions are mapped to short-range interactions.

Another point to notice is the constraint of negative semi-definiteness
of the off-diagonal elements, $H_{\sigma\sigma'}\le 0$.
This is necessary for $w_{\sigma\sigma'}$ to be positive as required for a transition matrix.
This condition excludes, for example, the interesting case of an antiferromagnetic fluctuation
term $\propto (\sum_i \sigma_i^x)^2$ with a positive coefficient in addition to the usual
transverse-field term with a negative coefficient in $\hat{H}$,
which has been shown to be effective to remove problematic first-order
quantum phase transitions \cite{SekiN,SeoaneN}.

It is possible to devise a quantum-to-classical mapping without the
above-mentioned negative semi-definiteness of off-diagonal elements \cite{Castel}.
However, in such a case, it is necessary to choose the eigenstates of $\hat{H}$ as
the basis of matrix representation to carry through the mapping,
which makes it difficult to interpret the resulting
classical Hamiltonian as an Ising model.

%%%%%%%%%%%%%%%%%%%%%%
\section{Time-dependent temperature}\label{sec:Tt}

If the temperature has explicit dependence on time as is the case in
most simulated annealing applications, the transition matrix also has time dependence.
This section is devoted to classical-to-quantum correspondence in such a case
\subsection{Classical to quantum mapping}

The master equation with time-dependent transition matrix is written as
\begin{equation}
\frac{d\hat{P}(t)}{dt}=\hat{W}(t)\hat{P}(t), \label{master_t-dependent}
\end{equation}
where $\hat{P}(t)$ is a vector with element $(\hat{P}(t))_{\sigma}=P_{\sigma}(t)$.
The corresponding quantum system is constructed as
\begin{equation}
\hat{H}(t)=-e^{\frac{1}{2}\beta (t)\hat{H}_0}\hat{W}(t)e^{-\frac{1}{2}\beta (t)\hat{H}_0}.
\end{equation}
If we introduce a wave function as
\begin{equation}
\hat{\phi}(t)=e^{\frac{1}{2}\beta (t)\hat{H}_0} \hat{P}(t),
\end{equation}
the master equation (\ref{master_t-dependent}) is rewritten as
\begin{equation}
-\frac{d\hat{\phi}(t)}{dt}=\left(\hat{H}(t)-\frac{1}{2}\dot{\beta}(t)\hat{H}_0\right)\hat{\phi}(t).
\label{i-Schrodinger}
\end{equation}
This is regarded as an imaginary-time Schr\"odinger equation:
If we rewrite the time as $t\to it$ in the time-derivative on the left-hand side,
the usual form of the Schr\"odinger equation results,
\begin{equation}
i\frac{d\hat{\phi}(t)}{dt}=\Big( \hat{H}(t)-\frac{1}{2}\dot{\beta}(t)\hat{H}_0\Big)\hat{\phi}(t).
\label{r-Schrodinger}
\end{equation}
Equations (\ref{i-Schrodinger}) and (\ref{r-Schrodinger}) show that an additional term
proportional to the time derivative of the inverse temperature is to be appended
to the quantum Hamiltonian to accommodate explicit time dependence of temperature
in the classical-to-quantum mapping.

\subsection{Convergence condition of simulated annealing}

Somma {\it et al.} \cite{SBO} discussed the convergence condition that the temperature
as a function of time, $T(t)$, should satisfy in simulated annealing for
the system to reach the ground state.
They used the classical-to-quantum mapping without explicit time dependence
of temperature as developed in Sec. \ref{sec:C_to_Q}, though in a slightly different
form as will be discussed below.
Then they applied the adiabatic theorem to the quantum system $\hat{H}$
and derived a result that is essentially equal to that of Geman and Geman \cite{GG},
$\beta (t)\propto \log t/pN$, where $p$ is an $\mathcal{O}(1)$ constant.
We discuss here a few problems in their analysis and show that their result turns out
to be justifiable by appropriately amending their argument.

First, the adiabatic theorem of quantum mechanics is derived from the
time-dependent Schr\"odinger equation, but they did not discuss explicitly
the relation between the original classical master equation, which
governs simulated annealing, and the Schr\"odinger equation.
Our result in Eq. (\ref{i-Schrodinger}) indicates that the master equation
is written as an imaginary-time Schr\"odinger equation, not the
usual real-time Schr\"odinger equation.
It has, nevertheless, been shown  \cite{MoritaN} that the adiabatic theorem
holds in the same form also for the imaginary-time Schr\"odinger equation,
which validates their analysis.

The second point concerns the additional term, $-\frac{1}{2}\dot{\beta}(t)\hat{H}_0$.
Somma {\it et al.} did not take this term into account.
However, according to their result, $\beta (t)\propto \log t/(pN)$,
the additional is inversely proportional to the system size and thus
can be neglected in the limit of large system size.
This serves as an {\it a posteriori} justification of their analysis
using only the $\hat{H}$ term.

The final comment is on the choice of the symmetric part of the transition matrix,
$w_{\sigma\sigma'}$, which they chose as $w_{\sigma\sigma'}=e^{-pN}$,
where $p\approx \max_j |H_j|$.
This is allowed as it does not violate the conditions
that the transition matrix should satisfy.
However, this choice of $w_{\sigma\sigma'}$ is different from the commonly-used heat-bath
and Metropolis methods,  which have explicit dependence on $\sigma$ and $\sigma'$.
This latter dependence is reflected in the dependence on $\sigma_{j-1}^z\sigma_{j+1}^z$
of the transverse-field term in Eqs. (\ref{hhat-hb}) and (\ref{hhat-Metropolis}).
Although it may happen that the final conclusion of Somma {\it et al.},
$\beta (t)\propto \log t/(pN)$,
does not depend upon the specific choice of the transition matrix, it is an interesting
problem to complete their analysis for more common types of  $w_{\sigma\sigma'}$.

%%%%%%%%%%%%%%%%%%%%%%
\section{Summary and conclusion}

We have analyzed the framework of classical-quantum correspondence of Castelnovo {\it et al.}
and have applied it to simulated annealing of the classical Ising model to study
its relation with quantum annealing using the transverse-field Ising model.
It has been shown that the eigenvalue spectrum is shared by the transition matrix
of the classical dynamics and the corresponding quantum Hamiltonian.
It then follows that the existence or absence of a phase transition and its order
are shared by the classical and quantum systems.
An important consequence is that simulated annealing of the classical Ising model
and quantum annealing by the corresponding transverse-field Ising model
have the same degree of efficiency as long as both are run
very slowly in the change of relevant parameters, that is, in quasi-equilibrium
classically and adiabatically in the quantum case.
Thus, simulated annealing and quantum annealing can be regarded as
equivalent if the transition matrix and the quantum Hamiltonian
are chosen to satisfy the key relation of Eq. (\ref{C_to_Q}).
The classical-to-quantum mapping has also been shown to provide an alternative
solution to the non-equilibrium dynamics of the one-dimensional Ising model.

The classical and quantum approaches, nevertheless, have an important difference
in the range of interactions in the Hamiltonians.
The classical-to-quantum mapping yields short-range interactions for the
quantum Hamiltonian if the range is short in the classical case,
but the converse is not true. The classical Hamiltonian generated from a quantum
system has in general very complicated many-body long-range interactions.
The range of interactions affects the efficiency in implementation of annealing,
and we may conclude that quantum annealing has a wider range of practical
usefulness.
This conclusion is reinforced by the restriction of the sign of matrix
elements of quantum Hamiltonian that can be mapped to classical dynamics.

System parameters such as the temperature are changed relatively rapidly
in practical applications of simulated annealing and quantum annealing.
We have formulated a classical-to-quantum mapping to cover such a case.
The Markovian dynamics has been shown to be mapped to an imaginary-time
Schr\"odinger dynamics with an additional term
proportional to the time-derivative of the inverse temperature.
This formulation would serve as a tool to analyze the performance
of rapid processes.

An overall conclusion is that simulated annealing and quantum annealing
share common aspects in their essential part in spite of the complete
difference of classical and quantum processes.
Quantum annealing, nevertheless, covers a wider range of
efficient implementation.

%%%%%%%%%%%%%%%%%%%%%
\begin{acknowledgments}
The work of HN was supported by JSPS KAKENHI Grant Number 26287086. 
HN and JT thank the Galileo Galilei Institute for Theoretical Physics
(Florence) for hospitality and INFN for partial support during the
completion of this work.
The work of SK was supported in part by the Office of the Director of
National Intelligence (ODNI), Intelligence Advanced Research Projects
Activity (IARPA), via IAA 145483; by the AFRL Information Directorate
under grant F4HBKC4162G001. The views and conclusions contained herein
are those of the authors and should not be interpreted as necessarily
representing the official policies or endorsements, either expressed
or implied, of ODNI, IARPA, AFRL, or the U.S. Government. 
The U.S. Government is authorized to reproduce and distribute reprints for
Governmental purpose notwithstanding any copyright annotation thereon.
\end{acknowledgments}

%%%%%%%%%%%%%%

\end{document}